# The effect of hydrogen on the magnetic properties of FeV superlattice

*M Elzain and M Al Barwani*
Department of Phyics, College of Science, Box 36, Sultan Qaboos University, Al Khod 123, Oman
Email: elzain@squ.edu.om

**Abstract** The electronic and magnetic structures of a hydrogenated and hydrogen free superlattice of 3 iron monolayers and 9 vanadium monolayers are studied using the first principle full-potential augmented-plane-wave method as implemented in WIEN2k package. The volume, the total energy and the magnetic moments of the system are studied versus the hydrogen positions at the octahedral sites within the superlattice and also versus the filling of the vanadium octahedral location by hydrogen atoms. It is found that the hydrogen locations at the interior of vanadium layer are energetically more favourable. The local Fe magnetic moment and the average magnetic moment per supercell are found to increase as the H position moves towards the Fe-V interface. On the other hand, the average magnetic moment per supercell is found to initially decrease up to filling by 3 H atoms and then increases afterwards. To our knowledge, this is the first reporting on the increase in the computed magnetic moment with hydrogenation. These trends of magnetic moments are attributed to the volume changes resulting from hydrogenation and not to electronic hydrogen-metal interaction.

## 1. Introduction

Inclusion of hydrogen into materials has both detrimental and beneficial effects. For example hydrogen embrittlement can cause a loss in ductility or load carrying ability at applied stresses well below the yield strength for metallic alloys[1]. On the other hand, materials with high hydrogen storage capacity are sought-for for use in clean energy[2]. Moreover, the interlayer exchange coupling (IEC) of magnetic superlattices that can be used in spintronics was found to be controllable by introduction of hydrogen in superlattice systems[3].

Iron and vanadium both possess cubic BCC structures at common temperatures and pressures with a lattice mismatch of about 5%. Iron and Vanadium form solid solution alloys over all concentration range[4]. Superlattices formed from Fe and V are known to exhibit perfect and sharp interfaces[5].

Hydrogen is readily and reversibly absorbed in vanadium metal. At low temperatures vanadium hydrides stabilize in the α and/or β phases. In the α phase the H atom are randomly distributed in the BCC structure, while in the ordered β phase H occupies the octahedral sites of the body-centred tetragonal structure[6]. In Fe/V superlattices the H atoms reside within the V layers[7] resulting in lattice expansion perpendicular to the layers planes[8]. Depletion regions of about two or three monolayers from the interfaces were observed[7].

The magnetization of iron-vanadium (FeV) alloys was found to decrease with increasing V concentration until it vanishes at about 70% at of V concentration for quenched samples and about 45% for slow cooled samples[9]. Addition of H to FeV alloys increases their magnetization[10]. Similarly, hydrogen inclusion in Fe/V superlattices was found to change the interlayer exchange coupling and to increase the magnetization of the Fe/V systems[11].

The electronic and magnetic properties of H in bulk V and FeV systems were studied using *ab initio* calculation. Andersson *et al.*[12] used the full potential linear muffin-tin-orbital and augmented plane-wave methods in the local density approximation to study the structural and electronic properties of $VH_x$. They concluded that the energy difference between the H occupation of octahedral and tetrahedral sites is too small to make any significant preference. It was also noted that hydrogenation of bulk vanadium induces weak antiferromagnetism. The location of H at interstitial sites in doped ultra-thin films of V was studied by Lebon *et al*[13] using the SIESTA code that employs linear combination of pseudoatomic orbitals. Hydrogen was found to prefer the tetrahedral sites for substitutional transition-metal dopants located on the left of V in the periodic table, while octahedral sites are preferred for dopants on the right. Meded and Mirbet[14] studied the H loading in superlattices of X/V, (X = Cr, Mo, Mn, Fe) using the projector augmented waves as employed in the VASP code. They found that H resides within the V layers out side the interface region with total energy trends following the accompanying volume expansion. The effect of H on the IEC and magnetic moment of a superlattice composed of three Fe layers and five V layers was studied by Ostanin et.al[15] using the full-potential linear-muffin-tin orbitals. The disappearance of the antiferromagnetic IEC for large H concentration was attributed to the decrease of the density of states (DOS) at the Fermi level. In addition, it was observed that hydrogen loading leads to the decrease of the interface magnetic moment of both Fe and V and hence to the decrease of the net magnetization contrary to the experimental findings.

Our main objective in this contribution is to study the effect of hydrogen on the net magnetic moment of Fe/V superlattices and to find out the sources of the gained magnetization on hydrogenation. We have used the full-potential linear-augmented-plane –wave (FP-LAPW) method as employed in the WIEN2k code[16] to calculate the electronic and magnetic properties of Fe/V superlattice composed mainly of three monolayers of Fe and nine monolayers of V. Our calculation, the details of which are presented in the following section, confirmed the previous findings. The changes in total energy are found to follow the trends in the changes in the local volume at the H location with interior interstitial positions being more favorable than the interstitial sites close to or at the Fe-V interface. The total magnetic moment of the unit cell was found to decrease initially with addition of H. With more H the moment increases to about 10% of the H free cell. These results are presented and discussed in section 3. Our conclusions are given in section 4.

## 2. Computational details

A superlattice with ideal interface consisting of three Fe monolayers and nine V monolayers is used in this study. According to the experimental observation of Hjorvarsson *et. al.*[3] and theoretical calculations[17], this superlattice lies within the region of the ferromagnetic IEC.

We have used a unit cell of H free 24 monolayers composed of consecutive 3 Fe monolayers and 9 V layers [$(Fe_3V_9)_2$] to calculate the energies of the ferromagnetically and antiferromagnetically interlayer coupled Fe layers. The antiferromagnetically coupled layers were found to be slightly more stable as compared to the ferromagnetic case. The calculation was repeated for the same unit cell with one H at the centre of the V layers. In this case the ferromagnetically coupled Fe layers were found to be slightly more stable. This agrees with the observation of Hjorvarsson *et al*[3] that addition of H changes the coupling to ferromagnetic. The required unit cell for further calculation is then reduced to half the initial unit cell. Thus a unit cell of 12 monolayers is used in all the calculations reported here.

The density functional L/APW+lo as implemented in WIEN2k package[16] is used to calculate the electronic and magnetic structures of FeV systems. In the L/APW+lo method the Kohn-Sham orbitals are expanded in terms of atomic orbitals inside the atomic muffin-tin (MT) sphere of radius $R_{MT}$ and in terms of plane waves in the interstitial regions. The Kohn-Sham equations were solved using Perdew-Burke-Ernerhof GGA approximation. Core and valence states were separated by an atomic energy of -7.0 Ry. In this case both the V 3s and the Fe 3s orbitals belong to the valence states. The calculations were carried out using an in-plane lattice constant of value 5.45 *a.u* obtained from the volume optimized $Fe_{0.5}V_{0.5}$ compound. The *c* lattice parameter is determined through volume optimization of the unit cell used. An MT radius of 1.86 *a.u* is used for Fe and V, while a radius of 0.9 *a.u* is used for H. For the valence electrons the potential and charge density are expanded in spherical harmonics up to L = 4, while the wavefunctions are expanded inside the MT sphere up to $l$ = 10 partial waves. Mixed basis are used depending on the partial wave $l$ with APW+lo being used for s, p and d valence orbitals. LAPW is used for the remaining $l$ values. For the plane wave expansion in the interstitial region we have used wavenumber cut off $K_{MAX}$ such that $R_{MT}K_{MAX} = 4.5$ and the potential and charge density are Fourier expanded with $G_{MAX} = 20$. The mesh size and the number of points in the Brillouin zone were tested for small tetragonal supercells using the convergence of the values of the electric field gradient and the total energy as indicators. A k sampling with a 16×16×1 Monkhost-Pack mesh is used in all calculations related to the superlattices. In all cases, the calculation is performed in the following order. The iteration for a fresh case is carried out until convergence in terms of total energy and charge density. Then the atomic positions along the z-direction are fully relaxed. This is followed by volume optimization. The calculation is then repeated using the relaxed atomic positions and new *c* parameter resulting from the volume optimization.

## 3. Results and discussions

Figure 1 shows the unit cell used in this study to calculate the electronic and magnetic properties of the hydrogenated Fe/V superlattice $Fe_3V_9$. We have performed calculations with H atoms at both Oz octahedral (see Schober and Wenzl, 1978) and tetrahedral interstitial sites. The figure shows H atoms at the Oz octahedral sites. The H atoms are indexed by numbers from -1 to 5 with H at the interface Fe being indexed 0. Hydrogen with index 5 lies at the centre of the V layer, while those with indices 1 and -1 lie at the V interface monolayer and the Fe interior monolayer respectively.

We have used the tetragonal space group P4mm (no. 99) to study the structure and the electronic and magnetic properties of the $Fe_3V_9$ hydrogen free superlattice and the effect of adding H atoms one atom at a time at the different Oz octahedral locations shown in figure 1.

The results of fully relaxed and volume optimized hydrogen free $Fe_3V_9$ superlattice show that the Fe-V interlayer distance is shorter than the Fe-Fe and V-V interlayer distances. This observation is also confirmed by a calculation on a hydrogen free $Fe_5V_9$ superlattice. The outer monolayer of Fe is found to have larger interlayer distance from the Fe monolayer below as compared to inner Fe interlayer distances. This may be considered as indication of stronger Fe-V interaction compared to Fe-Fe interaction.

### 3.1. The effect of single hydrogen atoms

The addition of H increases the unit cell volume particularly around its location. The change in volume of the cell where H resides from the corresponding H free case versus H location is shown in figure 2 together with the energy of the system with H in the same locations. We observe that the two quantities follow the same trend, where the maximum change in volume and the highest energy occur when H is at the Fe interface monolayer. The least change in volume and the lowest energy are observed when H is at the V central monolayer. This agrees with the observation of Meded and Mirbt[14] for $V_6Fe_6$ superlattice.

The presence of hydrogen influences the magnetic moment of the individual Fe and V atoms. However, its net effect can be assessed by considering the net magnetic moment per unit cell. Table 1 shows the net magnetic moment per unit cell per Fe atom with one H atom at the indicated locations together with the local magnetic moments at the individual atoms. The table consists of 5 sets of results. The first column under $Fe_3V_9$ shows the results of the fully relaxed and volume optimized hydrogen free $Fe_3V_9$ superlattice. The first row gives the average magnetic moment per unit cell per Fe atom for different H locations. Using the same relaxed atomic positions and optimized volumes obtained for the different H locations, calculations were repeated without H. The resulting average magnetic moments per unit cell per Fe atom are shown in the second row. We note that the results in the two rows are comparable. This may signify that the resulting changes in the magnetic moment could be attributed to volume effect and not to H-metal electronic interactions. In both cases, the maximum increase in the moment (as compared to the H free $Fe_3V_9$ moment) is obtained for H location at interface Fe monolayer. The net moment monotonically decreases for H locations away from the interface reaching its minimum when H is at the centre of the V layer.

The following set of data in table 1 gives the local magnetic moment within the muffin-tin sphere at the interface Fe monolayer (Fe (0)), the central Fe monolayer ( Fe (-1)), and the V interface monolayer (V (1)) for different H locations. The last set in this table gives the moments at the same sites when the calculation is repeated using the same relaxed atomic positions and the same optimized volumes as obtained for different H location but calculated without including H. The effect of hydrogenation is deduced from comparison to the hydrogen free $Fe_3V_9$ superlattice results shown in the first column. When H is located at the V interior layers no significant changes in the magnetic moments of Fe and V at the interface layers are observed. Clear changes arise when H is at or close to the interface. For H at the interface Fe monolayer, the magnetic moment of the interface Fe atom increases considerably from 1.57 $\mu_B$ to 2.03 $\mu_B$, whereas the moments of the Fe atom at the central layer and the V at the interface slightly decrease. The Fe moment at the interface Fe monolayer increases to 1.86 $\mu_B$ when H is located at the V interface. The same trends are reflected by the corresponding H free cases. However, the increase of the interface Fe moment is more prominent and in addition the interface V moment shows an increase in magnitude.

It is well documented that the magnetic moment of the Fe atoms at the interface in the Fe/V superlattices is reduced due to their interaction with the neighboring V atoms, while the magnetic moment of the next inner Fe monolayer is increased [18]. The results of the hydrogen free $Fe_3V_9$ superlattice shown in the first column of table 1 agree with the reported findings. Addition of H in the proximity of the interface increases the interface Fe moment. We argue that this could be attributed to volume effect. Consider the case where H atom is at the 0 location. Here the H atom neighbors both the Fe atoms in central monolayer and the V atom at the interface. We find that hydrogen interacts more strongly with Fe compared to its interaction with V as is clearly exhibited by the charge density maps of spin-up and spin-down electrons depicted in figure 3. The relaxed H-Fe and H-V distances in this case are found to be 3.11 a.u and 3.37 a.u respectively. The H interacts more strongly with the spin-up electrons of Fe pushing their energy upward and hence converting electrons from spin-up levels to the spin-down levels resulting in the reduction of the magnetic moment of the Fe atom. The opposite happens in the case of V since the spin-down levels are lower in energy than the spin-up levels. The presence of H at the 0 location increases the interlayer distance between Fe and V at the interface from 2.62 a.u to 3.30 a.u leading to reduction in Fe-V interaction and consequently to the observed increase in the magnetic moment of the Fe atoms at the interface. The set of data at the bottom of table 1 for the corresponding H free locations confirm this interpretation. The observed increase in Fe and V moments in the latter case is due to the absence of H-metal interaction.

### 3.2. The effect of multi hydrogen atoms

From the energetic point, the first hydrogen atom entering the $Fe_3V_9$ superlattice will reside at the centre of the V layer. The question we want to address is how the filling of the interstitial sites proceeds on further hydrogenation. In addition to one H atom at the central V monolayer we have added H atoms one at a time at the other Oz sites forming supercells with two H atoms. The trends in their energies and magnetic moments were found to be identical to the case of single H atom supercells discussed above. Furthermore, we considered the occupation of two H atoms of different tetrahedral sites. In all cases, when comparing the energies of supercells with H atoms at tetrahedral sites with the corresponding energies of supercells with atoms at the Oz sites in the same proximity, it turns out that the cases where H atoms are at the Oz sites are more favorable.

Henceforth, we assume that on hydrogenation of the Fe/V superlattices H occupies the Oz sites starting with the position at the central V monolayer and proceeds successively towards the interface. Since it was experimentally confirmed that a depletion region of thickness of 2-3 monolayers exists at the interface, we considered filling by H up to location 3 only. The question of what happens on further hydrogenation is not addressed. We show in table 2 the average magnetic moments per Fe atom and the local moments versus H filling of the Oz sites. We note that the average moment per Fe atom initially decreases with addition of H up to filling with three H atoms and then starts increasing. At the same time we found that the volume of the corresponding supercells increases monotonically with increasing number of H atoms. Figure 4 shows a plot of the supercell volume and the average magnetic moment per Fe atom versus the H/V fraction. The relative perpendicular lattice expansion due to H loading and the magnetization of $Fe_3V_{11}$ superlattice as measured by Labergerie et al[11] exhibit a somewhat similar logarithmic trends with H content.

The corresponding average moment for the cases with the same relaxed positions and the same volume but without H exhibit an almost steady increase. Furthermore, the local Fe moments with and without H also steadily increase. We found that the difference in the trends of the magnetic moments between the hydrogenated and the hydrogen free cases results from the small contribution of the inner V monolayers. Although these contributions are very small but on addition of H the polarization of some of the vanadium atoms changes sign to produce the small difference between the trends of the moments. The origin of the increase in the average and local moments can also be attributed to the volume effect. The interlayer distances at interface hardly change with introduction of H at the interior Oz locations in the V layer. However, the inner interlayer spacing expands reaching values of order 3.6 a.u. and leads to large supercell volumes. Through indirect interaction with the inner V atoms, the Fe atoms at the interface gain more moment. To test this we computed the Fe moment in a smaller supercell consisting of one Fe monolayer and five V monolayers without hydrogen. We kept the interlayer distance of the Fe monolayer to its first and second neighboring V monolayers on both sides fixed, while increasing the interlayer distance of the outer V monolayer. As shown in figure 5, the magnetic moment of the Fe atom remains zero for interlayer distance of order 3.5 a.u. or less. For higher interlayer distances the Fe atom starts gaining magnetic moment. Further support to the effect of volume in increasing the local magnetic moment is seen from the table 2. We note that the local magnetic moments calculated for the supercells with and without H follow the same trends and their values are in close comparison. Thus the effect of the presence of H in the V layer is to increase the volume of the supercell without considerable electronic structure consequences on the Fe atoms.

## 4. Conclusion

The electronic and magnetic structures of $Fe_3V_9$ superlattice with and without hydrogen were calculated using the *ab initio* FP-LAPW method. Hydrogen was included at various octahedral Oz sites varying from the centre of the Fe layer to the centre of the V layer. From the comparison of total energies of the supercells with one hydrogen atom at the different octahedral sites it is concluded that H favors the interior Oz locations of the V layer. The total energy was found to follow the trends of the resulting change in volume due to inclusion of hydrogen in agreement with previously reported results. Hydrogen was found to bond more strongly to Fe and less so to V atoms. The average magnetic moment of the fully-relaxed and volume-optimized supercell with single H atoms at different locations was found to increase as H position moves from the centre of the V layer towards the Fe-V interface. The same trends were obtained for the H free supercell using the same atomic positions and volume of the corresponding supercell with hydrogen. It is concluded that the increase in magnetic moment at the interface is linked to the interlayer expansion due to hydrogenation.

The effect of filling of Oz sites with H atoms on the magnetic moment was also studied. The average magnetic moment per fully-relaxed and volume-optimized supercell was found to initially decrease by addition of hydrogen up to 3 atoms per supercell. The average magnetic increases afterwards. On the other hand, the supercell volume was found to be monotonically increasing. The average magnetic moment of the corresponding supercell without H but with the same atomic positions and volume was found to increase monotonically. Furthermore the local magnetic moment of the Fe atoms was found to exhibit the same trends of the average moment as in the latter case. This suggests volume expansion as the major contributor to the changes of the magnetic moment.

**References**

1. N. Eliaz, A. Shachar, B. Tal and D. Eliezer, Eng. Failure Anal **9**, 167 (2002).
2. B. Sakintuna, F. Lamari-Darkrim and M. Hirscher, Inter. J. Hydr. Energy **32**, 1121 (2007).
3. B. Hjorvarsson, J. A. Dura, P. Isberg, T. Watanabe, T. J. Udovic, G. Andersson and C. F. Majkrzak, Phys. Rev. Lett. **79**, 901 (1997).
4. J. M. Sanchez, M. C. Cadeville, V. Pierron-Bohnes and G. Inden, Phys. Rev. B **54**, 8958 (1996).
5. G. Andersson, E. Nordstrom and R. Wappling, Eur. Phys. Lett. **60**, 731 (2002).
6. T. Schober and H. Wenzl, in *Hydrgen in Metals* edited by A. Alefeld (Springer-Verlag, Berlin, 1978), Vol. II pp. 11.
7. A. Remhof, G. Nowak, H. Zabel, M. Bjorck, M. Parnaste, B. Hjorvarsson and V. Uzdin, Eur. Phys. Lett. **79**, 37003 (2007).
8. G. K. Pálsson, V. Kapaklis, J. A. Dura, J. Jacob, S. Jayanetti, A. R. Rennie and B. Hjörvarsson, Phys. Rev. B **82**, 245424 (2010).
9. J. C. Krause, J. Schaf, M. I. da Costa, Jr and C. Paduani, Phys. Rev. B **61**, 6196 (2000).
10. A. Obermann, W. Wanzl, M. Mahnig and E. Wicke, J. Less Comm. Metals **49**, 75 (1976).
11. D. Labergerie, K. Westerholt, H. Zabel and B. Hjorvarsson, J. Magnet. Magnet. Mater. **225**, 373 (2001).
12. P. Andersson, L. Fast, L. Nordstrom, B. Johansson and O. Eriksson, Phys. Rev B **58**, 5230 (1998).
13. A. Lebon, A. Vega and A. Mokrani, Phys. Rev B **81**, 094110 (2010).
14. V. Meded and S. Mirbit, Phys. Rev B **71**, 024207 (2005).
15. S. Ostanin, V. M. Uzdin, C. Demangeat, J. M. Wills, M. Alouani and H. Dreyssé, Phys. Rev B **61**, 4870 (2000).
16. P. Blaha, K. Schwarz, G. K. H. Madsen, D. Kvasnicka and J. Luitz, in *WIEN2k: an augmented plane wave + local orbitals program for calculating crystal properties)* edited by K. Schwarz (Technische Universitat Wien, Austria, Vienna, 2009).
17. B. Skubic, E. Holmstrom, A. Bergman and O. Eriksson, Phys. Rev. B **77**, 144408 (2008).
18. V. Uzdin, K. Westerholt, F. Zabel and B. Hjorvarsson, Phys. Rev. B **68**, 214407 (2003).

Table 1. The average magnetic moments per supercell in Bohr Magnetons (first row), versus the position of one hydrogen atom at the Oz locations given by, the indices from -1 to 5. The second row gives the corresponding average magnetic moments for H free supercells calculated using the same relaxed atomic positions and volume optimized cases as in the first row. The third and fourth set of data give the local magnetic moments at the interface Fe (0), the interior Fe (-1) and the interface V (1) for hydrogenated and the corresponding H free cases. The first column gives the average and local magnetic moment for the fully-relaxed and volume optimized supercell without hydrogen.

| Location | Fe$_3$V$_9$ | -1 | 0 | 1 | 2 | 3 | 4 | 5 |
|---|---|---|---|---|---|---|---|---|
| | | | Average moment (with H) | | | | | |
| | | 1.60 | 1.68 | 1.65 | 1.58 | 1.59 | 1.56 | 1.51 |
| | | | Average moment (without H) | | | | | |
| | 1.55 | 1.63 | 1.70 | 1.65 | 1.56 | 1.57 | 1.58 | 1.54 |
| | | | Local moment (with H) | | | | | |
| Fe (0) | | 1.57 | 2.03 | 1.86 | 1.63 | 1.63 | 1.59 | 1.55 |
| Fe (-1) | | 2.59 | 2.42 | 2.54 | .58 | 2.54 | 2.56 | 2.54 |
| V (1) | | -0.37 | -0.28 | -0.32 | -0.37 | -0.30 | -0.32 | -0.33 |
| | | | Local moment (without H) | | | | | |
| Fe (0) | 1.57 | 1.61 | 2.28 | 1.97 | 1.58 | 1.59 | 1.57 | 1.54 |
| Fe (-1) | 2.55 | 2.79 | 2.58 | 2.51 | 2.58 | 2.52 | 2.55 | 2.55 |
| V (1) | -0.32 | -0.40 | -0.53 | -0.44 | -0.35 | -0.31 | -0.32 | -0.32 |

Table 2. The average magnetic moments per supercell in Bohr Magnetons (first row), versus the fraction of hydrogen atoms at the Oz locations (H/V). The second row gives the corresponding average magnetic moments for H free supercells calculated using the same relaxed atomic positions and volume optimized cases as in the first row. The third and fourth set of data give the local magnetic moments at the interface Fe (0), the interior Fe (-1) and the interface V (1) for hydrogenated and the corresponding H free cases.

| H/V | 0 | 1/9 | 2/9 | 3/9 | 4/9 | 5/9 |
|---|---|---|---|---|---|---|
| Average moment (with H) | | | | | | |
|  | 1.55 | 1.51 | 1.53 | 1.51 | 1.62 | 1.70 |
| Average moment (without H) | | | | | | |
|  | 1.55 | 1.54 | 1.57 | 1.57 | 1.61 | 1.63 |
| Local moment (with H) | | | | | | |
| Fe (0) | 1.57 | 1.55 | 1.59 | 1.61 | 1.66 | 1.67 |
| Fe (-1) | 2.56 | 2.54 | 2.58 | 2.60 | 2.59 | 2.63 |
| V (1) | -0.32 | -0.33 | -0.34 | -0.35 | -0.33 | -0.33 |
| Local moment (without H) | | | | | | |
| Fe (0) | 1.57 | 1.54 | 1.59 | 1.58 | 1.62 | 1.63 |
| Fe (-1) | 2.56 | 2.55 | 2.58 | 2.59 | 2.58 | 2.60 |
| V (1) | -0.32 | -0.32 | -0.33 | -0.33 | -0.33 | -0.34 |

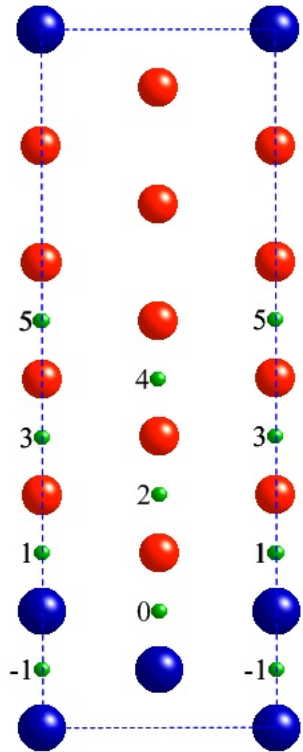

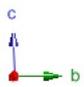

Figure 1. The unit cell of the $Fe_3V_9$ superlattice. Dark (blue) large balls represent Fe and light (red) large balls represent V. The small (green) balls represent H at the (Oz) octahedral interstitial sites. The indices give the hydrogen locations within the superlattice.

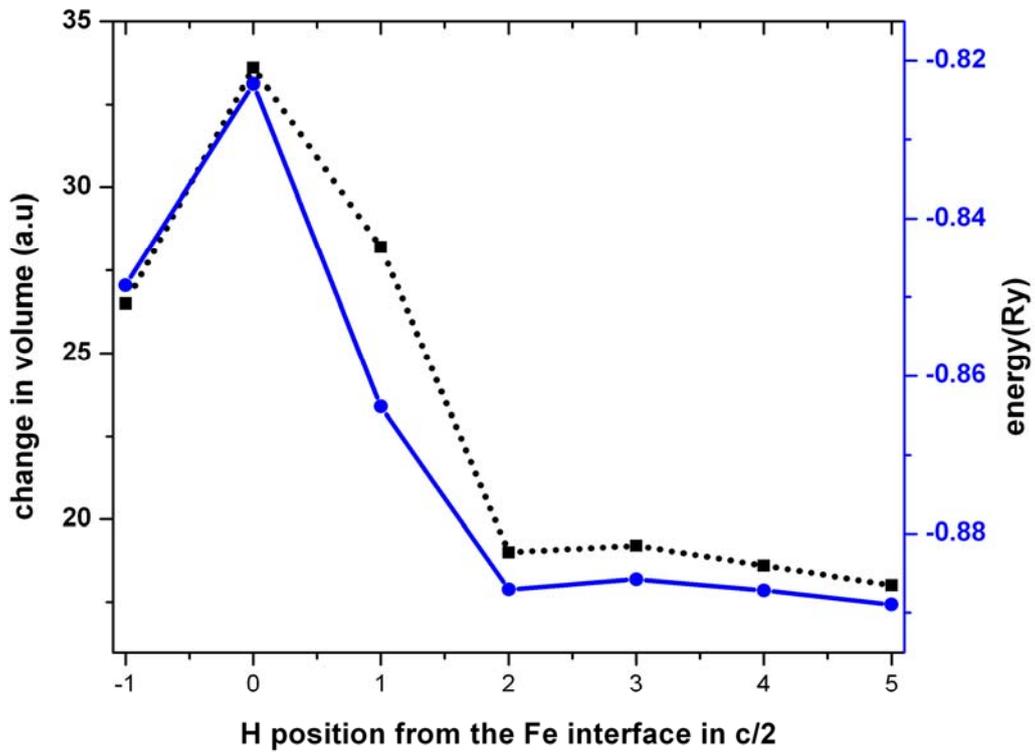

Figure 2. The change in volume between the fully relaxed and volume-optimized cell containing one hydrogen atom and the fully relaxed and volume-optimized hydrogen free case (continuous, blue curve) and the corresponding relative total energy (dotted, black curve) versus the hydrogen positions at the Oz octahedral sites.

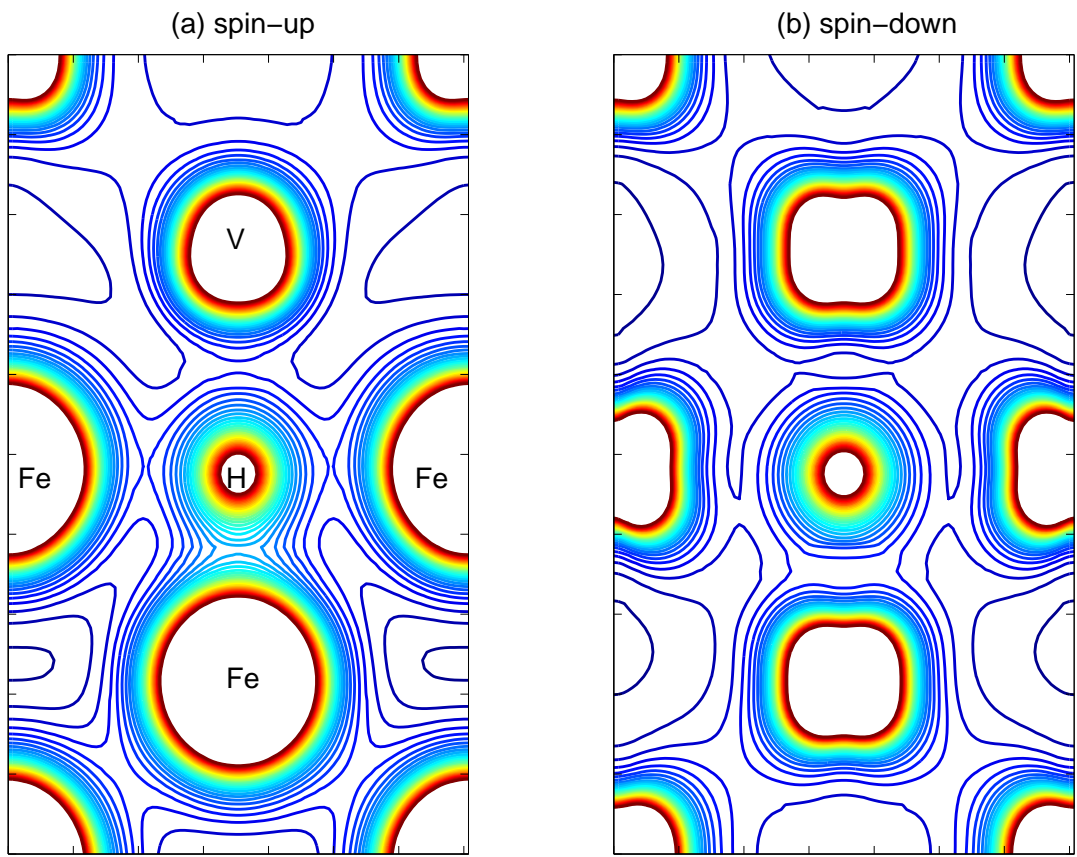

Figure 3. The charge density maps on the (110) surface of the supercell with hydrogen at the Oz site of interface iron monolayer Fe (0) for spin-up (a) and spin-down (b) electrons. The atoms of the spin-down case (b) are identical to those of the spin-up case (a)

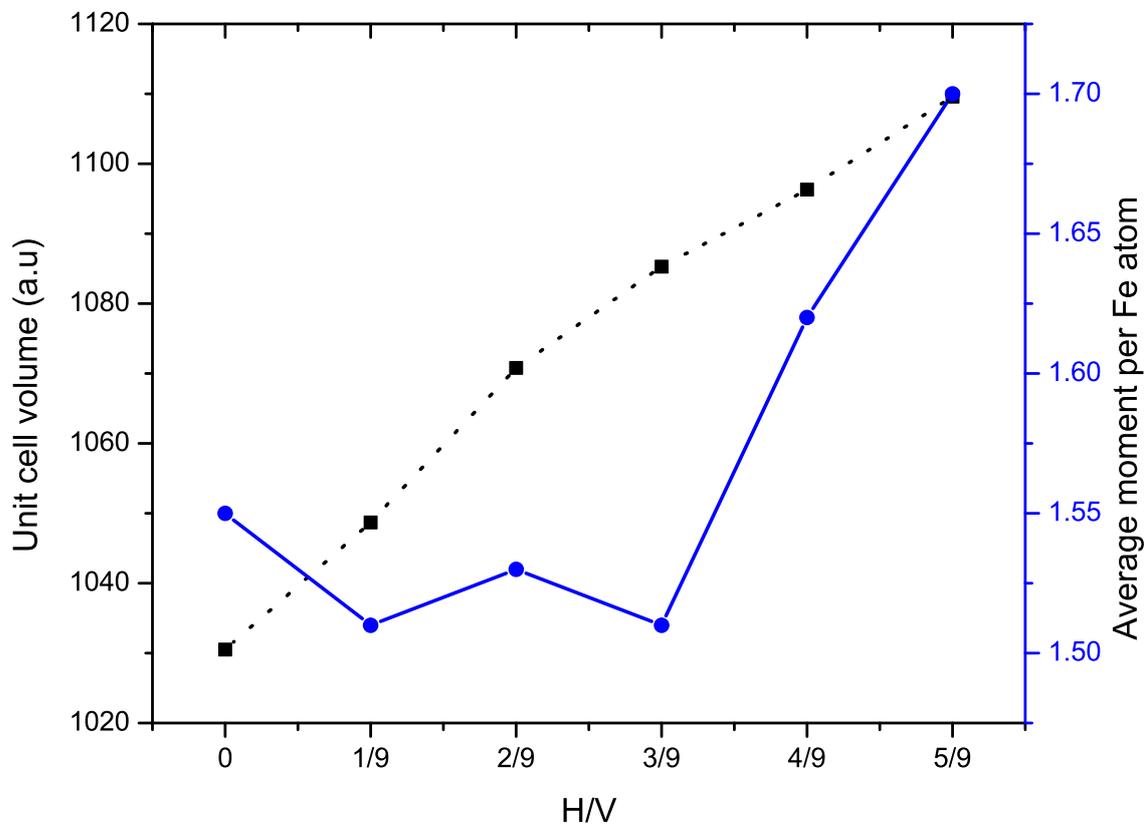

Figure 4. The volume of the supercell (dashed, black curve) and the average magnetic moment per supercell (continuous, blue curve) versus the fraction of hydrogen (H/V) at the Oz locations.

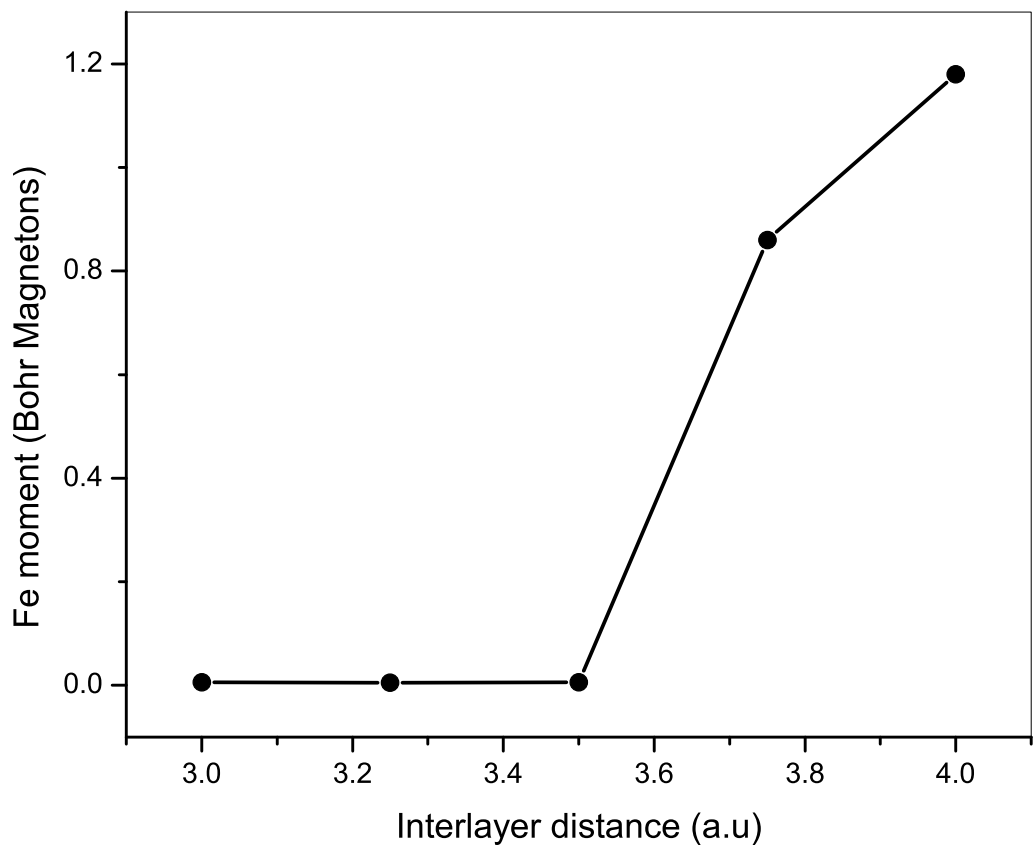

Figure 5. The local magnetic moment of Fe atoms in the superlattice $Fe_1V_5$ at fixed interlayer distances of Fe monolayer to its nearest and next-nearest neighbouring V monolayers versus the interlayer spacing of the outer V monolayer.